\documentclass[aps,prl,reprint,superscriptaddress,floatfix]{revtex4-2}
\usepackage{graphicx} % Required for inserting images
\usepackage[total={6.5in, 9in}]{geometry}
\usepackage[normalem]{ulem}
\usepackage{xcolor}
\usepackage{soul}
\usepackage{amsmath}
\usepackage{amssymb}
\usepackage{textcomp, gensymb}

\newcommand\raulst{\bgroup\markoverwith{\textcolor{magenta}{\rule[0.5ex]{2pt}{0.4pt}}}\ULon}

\begin{document}

%TC:ignore

\title{Bubbling in Oscillator Networks}
\author{Giulio Tirabassi}
\thanks{These authors contributed equally to this work.}
\affiliation{Universitat de Girona, Departament de Informàtica, Matemàtica Aplicada i Estadística, Universitat de Girona, Carrer de la Universitat de Girona 6, Girona 17003, Spain.
}

\author{Raul de Palma Aristides}
\thanks{These authors contributed equally to this work.}
\affiliation{Pompeu Fabra University, Department of Medicine and Life Sciences, Carrer del Doctor Aiguader 88, Barcelona 08003, Spain.
}% 

\author{Cristina Masoller}%
\affiliation{Universitat Politècnica de Catalunya, Departament de Fisica, Rambla Sant Nebridi 22, Terrassa 08222, Barcelona, Spain.
}%

\author{Daniel J. Gauthier}
\affiliation{ResCon Technologies, LLC, 1275 Kinnear Rd., Suite 239, Columbus, Ohio 43212, USA}

\date{\today}

\begin{abstract}
A network of coupled time-varying systems, where individual nodes are interconnected through links, is a modeling framework widely used by many disciplines. For identical nodes displaying a complex behavior known as chaos, clusters of nodes or the entire network can synchronize for a range of coupling strengths.  Here, we demonstrate that small differences in the nodes give rise to desynchronization events,  known as bubbling, in regimes where synchronization is expected.  Thus, small unit heterogeneity in all real systems has an unexpected and outsized effect on the network dynamics.  We present a theoretical analysis of bubbling in chaotic oscillator networks and predict when bubble-free behavior is expected.  Our work demonstrates that the domain of network synchronization is much smaller than expected and is replaced by epochs of synchronization interspersed with extreme events. Our findings have important implications for real-world systems where synchronized behavior is crucial for system functionality.
\end{abstract}

\maketitle
%TC:endignore

\section{Introduction}

\begin{figure*}[tb] %%% fig 1
    \centering
    \includegraphics[width=0.9\linewidth]{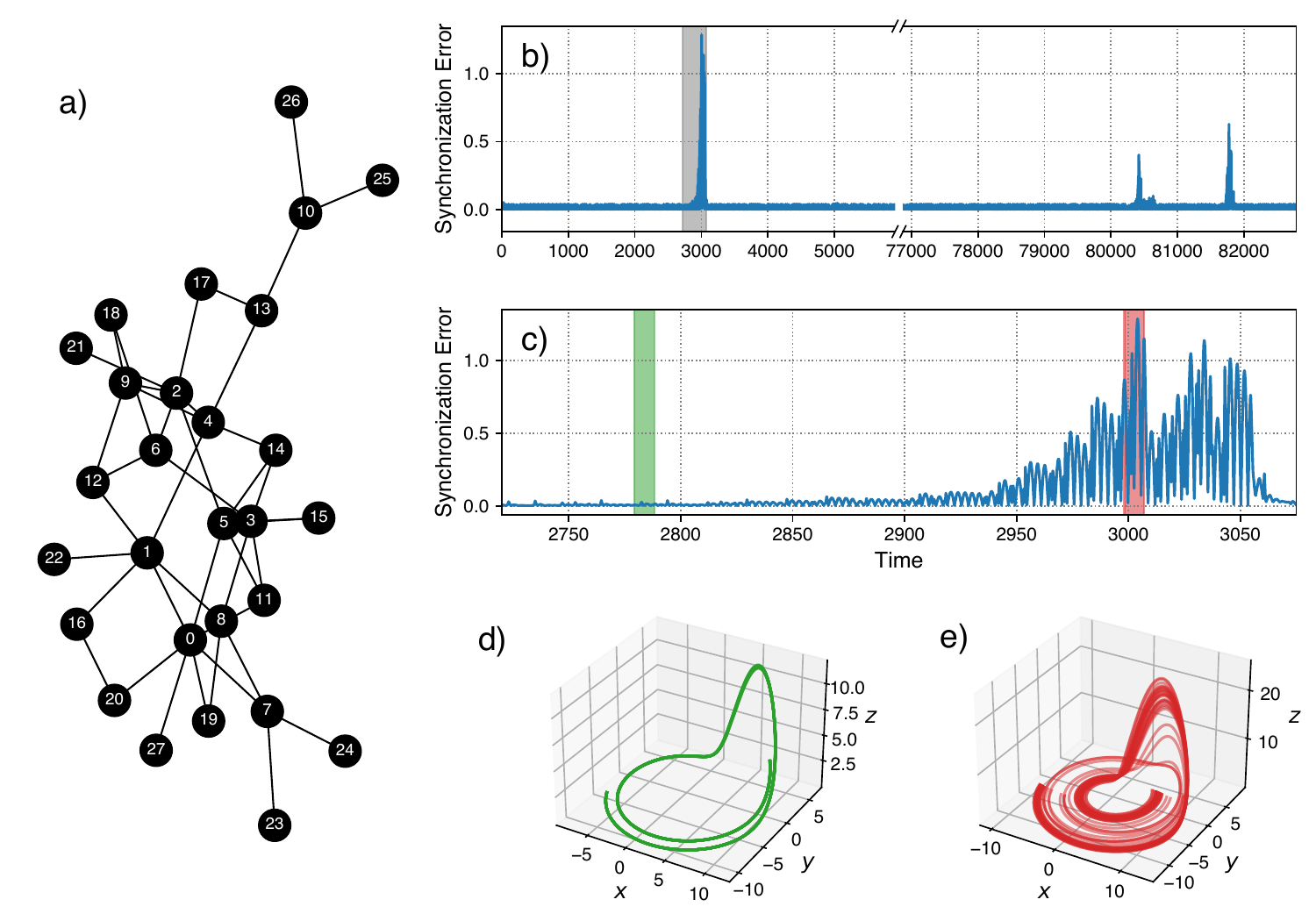}
    \caption{\textbf{Attractor bubbling in complex oscillator networks.} a) Topology of a $N=28$ node network of chaotic Rössler oscillators with undirected links. b) Bubbling events appear as rare and large fluctuations in the temporal evolution of the synchronization error.  The break in the horizontal axis allows us to display three of these rare extreme events. c) Temporal evolution of the synchronization error during the bubble occurring at $t \approx 3,000$, indicated by the gray rectangle in b). Trajectories of the individual oscillators during the d) synchronized precursor (green rectangle in c)) and e) full development (red rectangle in c)) of the event. The coupling parameter is $\sigma=0.1$ and the other parameters are given in \textit{Methods}.  An animation of a bubbling event is given the \textit{Supplementary Information.}
    }
    \label{fig:storyboard}
\end{figure*}

A network consists of nodes that accept input signals, process the information, and pass the output to other nodes in the network over links. Figure \ref{fig:storyboard}a) illustrates a network, where the circles represent the nodes and the lines connecting the circles represent bi-directional links. Often, the node outputs are nonlinear functions of the inputs and have internal time-dependent behavior, and the links are linear and have different connection weights.  The network concept is a simplified abstraction of complex systems studied in various disciplines. Yet, conclusions drawn from studies of generic network dynamics often have universal application \cite{Newman2018}.

One key challenge is identifying the conditions when the dynamics of the network's nodes synchronize \cite{misha,arenas} because it represents a coherent response, which is often important for system functioning (see, for example, Refs.~\cite{motter2013, Schafer2018}). One approach for studying network synchronization is to assume that the network has no external inputs and that the nodes are identical dynamical systems.  A key finding from research in the early 80's is that, under these assumptions, the network can fully synchronize and that the dynamics of each network node are restricted to a limited region of the phase space, known as the synchronization manifold \cite{Fujisaka1983}.

The stability of the synchronized state is determined by studying the network's response to perturbations that are transverse to the synchronization manifold. A popular tool used to analyze the asymptotic stability of the synchronized state is the master stability function (MSF)~\cite{Pecora1998}, which assumes that the network nodes are linearly-coupled identical dynamical systems. Under this assumption, the stability of the synchronized state of the network is the same as the stability of the synchronized state of two nodes, where the effect of the network topology is encoded in the coupling matrix that describes the interaction between the nodes (we present an overview of the MSF approach in {\it Methods}).  

In the MSF approach, the largest transverse Lyapunov exponent, $\lambda_{max}^{\perp}$, determines the linear asymptotic stability of the synchronized state against transverse perturbations \cite{Pikovsky_2016}. 
Stability is predicted when $\lambda_{max}^{\perp}<0$.  Often, the coupling parameter $\sigma$ is used to control the synchronized state stability with a threshold value $\sigma_{th}$ predicted by the MSF (see Eq. (\ref{eq:sigma_definition}) in {\it Methods}).  The MSF has been widely used to study network synchronization \cite{chen2003,diBernardo2009, Sorrentino2012,dsouza2019,mulas2020,perc2021,Zhang2021a,Zhang2021b, Fahimipour2022, Clusella2023, Lizier2023, Smith2024, Nazerian2024} because it provides a simple stability criterion.  However, it is only one of several possible criteria \cite{Pecora1998} and, as discussed below, it is a necessary but not sufficient condition for high-quality synchronization.

The MSF method has been generalized to predict the synchronization of a network of similar but not identical oscillators \cite{Restrepo2004} and when the sums of the entries in the rows of the coupling matrix slightly deviate from zero \cite{sorrentino2011}.  Here, the dynamics of the nodes are close to the synchronization manifold if the heterogeneity of the nodes is small enough \cite{Ott_pre_1996,Sun2009}. 

Another generalization predicts the appearance of cluster synchronization, where a subset of nodes synchronize, while others do not, or there are other subsets each of which is synchronized to distinct dynamics \cite{Zhou2006, Pecora2013, Sorrentino2016, Schaub2016, Bayani2024}. Furthermore, propagation delays along the links \cite{DHuys2008,Dahms2012} or large node heterogeneity \cite{Nazerian2023} can be incorporated into the MSF stability analysis.   

These criteria do not exclude the possibility of brief but large-scale desynchronization events, which may be catastrophic in some networked systems such as electric power networks \cite{Sajadi2022}.  During such an event, which can occur even when $\lambda_{max}^{\perp}<0$, there is a departure from the synchronization manifold followed by a return to the manifold. 

\textit{Attractor bubbling} refers to these desynchronization events  \cite{Ashwin1994,Heagy1995,Gauthier1996,Ott_pre_1996,Blakely2000a,Blakely2000b,Yanchuck2001,Restrepo2004,Cavalcante2013} and usually occurs for coupling strengths near the MSF stability boundary ($\sigma_{th}$ for which $\lambda_{max}^{\perp}=0$), and for stronger coupling, for which $\lambda_{max}^{\perp}<0$. In this coupling region, bubbling can also occur if the nodes are identical but affected by noise \cite{Ashwin1994,Ott_pre_1996,Venkataramani1997,Restrepo2004}.

Examples of bubbling events in a network of chaotic R\"{o}ssler oscillators with tiny parameter mismatches ($<$ 0.5\%) are shown in Fig.~\ref{fig:storyboard} (the model and parameters are presented in \textit{Methods}). Here, the coupling parameter $\sigma$ is within the range of coupling strengths where the MSF predicts asymptotically stability. However, we see in Fig.~\ref{fig:storyboard}b) that long epochs of small synchronization error (SE) are interrupted by large desynchronization events. The error is defined as
\begin{equation}
\mathrm{SE}(t)=\sum_n \left|\mathbf{x}^n(t) - \bar{\mathbf{x}}(t)\right|/N, 
\end{equation}
where $\mathbf{x}^n$ denotes the vector of variables of the $n^{th}$ node and $\bar{\mathbf{x}}(t)=\sum_n \mathbf{x}^n(t)/N$.

Figure~\ref{fig:storyboard}c) shows the detail of a bubble, where the SE is initially small, but the trajectories of the nodes separate as the event proceeds.  The distances between pairs of oscillators can be comparable to the size of the attractor during a fully evolved bubble.  High-quality synchronization, where the node trajectories stay close to the synchronization manifold, can be observed, but for $\sigma \gg \sigma_{th}$.

This paper shows that bubbling is pervasive in networks of R\"{o}ssler chaotic oscillators, appearing over a wide range of coupling strengths.  For network topologies where cluster synchronization is expected, we find that the hierarchical sequence of cluster formation predicted by the MSF is preserved \cite{Bayani2024} but the regions of coupling strength where bubble-free synchronization occurs are considerably smaller. 

To understand these observations, we systematically study five stability criteria based on measures of transverse stability of the synchronization manifold of a network of identical oscillators.  We find that none of these criteria predict the bubbling domain.  We therefore introduce a new stability criterion that include these measures and the duration of the time interval that the system resides in the transversely unstable regions of the synchronization manifold.  One measure -- based on the finite-time transverse Lyapunov exponents and the duration of the averaging window (defined below) -- is the only one that correctly indicates the bubbling domain.

While the presence of bubbling has been known for decades for two coupled chaotic oscillators, it is largely ignored by the community of researchers studying dynamics of oscillator networks.  We purposely study networks of chaotic R\"{o}ssler oscillators because it is often the system of choice in previous studies, which allows us to make a direct comparison to their work.  A primary contribution of our paper is to emphasize that bubbling is pervasive.  It is important to point out this issue because bubbling may have catastrophic consequences for natural or human-designed networks if synchronized behavior is expected.

\section{Results}

In this section, we systematically evaluate the criteria for predicting the domains of high-quality synchronization and bubbling.  A boxed equation highlights each criterion.  We begin by defining the relevant concepts and notation.  We end with results for different network sizes and cluster synchronization.

The transverse Lyapunov exponents quantify the exponential separation of trajectories over time when starting from two different initial conditions close to the synchronization manifold (see \textit{Methods}). A classic way to estimate the largest exponent from discretely-sampled data (sample period $dt$)  
is to observe the trajectories briefly, estimate the growth or decay rate, reset the separation, and repeat many times around the synchronization manifold \cite{Abarbanel1992,holger}. At each step, the distance $\delta \mathbf{x}$ between the trajectories scales as
\begin{equation}
\delta \mathbf{x}=\exp({\lambda_i^{\perp} dt}),
\end{equation}
where the \textit{transverse local Lyapunov exponent} (LLE) \cite{Benettin1980} is denoted by $\lambda_i^{\perp}$. It is well known that local Lyapunov exponents have large variation on a chaotic attractor \cite{Abarbanel1991} and we show below that this is also the case for the synchronization manifold.

The largest \textit{finite-time transverse Lyapunov exponents} \cite{Abarbanel1991,Abarbanel1992} is the average of the LLEs over a time interval $\tau_m = m~dt$ and is given by 
\begin{equation}
    \bar{\lambda}_m^{\perp} = \frac{1}{m}\sum_{i=1}^m \lambda_i^{\perp}. \label{eq:finite_time_def}
\end{equation}
The largest transverse Lyapunov exponent, mentioned in the Introduction, is given by
\begin{equation}
    \lambda_{max}^{\perp} = \lim_{m\rightarrow \infty} \bar{\lambda}_m^{\perp}. \label{eq:Lyapunov_def}
\end{equation}

Transverse stability can also be assessed by considering invariant sets embedded in the attractor on the synchronization manifold. For the R\"{o}ssler oscillator considered here, the invariant sets related to bubbling are unstable periodic orbits (UPOs) \cite{Heagy1995}. The trajectories of the oscillators visit the $j^{th}$ UPO 
for a typical time given by \cite{Ott_pre_1996}
\begin{equation}
    \tau_j^{UPO} = 1/\lambda_j^{\parallel, UPO}, \label{eq:UPO_time}
\end{equation}
where $\lambda_j^{\parallel, UPO}$ is the largest Lyapunov exponent associated with this UPO on the synchronization manifold (indicated by $\parallel$). Here, $\lambda_j^{\parallel, UPO}$ is found by forcing the trajectories to stay in the neighborhood of the $j^{th}$ UPO and the UPOs can be found by searching for near-repeated points on the attractor.  After a certain time $\tau_{UPO}$ the trajectory leaves the UPO and visits another UPO, a process that repeats again and again.  These orbits are unstable and their transverse stability is determined by their associated transverse Lyapunov exponents $\lambda_j^{\perp,UPO}.$

To quantify the quality of the synchronization, we analyze the \textit{synchronization error}, whose temporal evolution during a typical bubble is shown in Figs.~\ref{fig:storyboard}b) and c).

\textbf{Criterion \#1: The Master Stability Function}  

As stated in the previous section, the MSF predicts that all oscillators in the network synchronize when
\begin{equation}
    \boxed{~~\lambda_{max}^{\perp} < 0.~~} \label{eq:MSF_criterion}
\end{equation}

The typical method to assess whether the MSF correctly predicts synchronized behavior is to observe the network dynamics over a long time for each $\sigma$.  %, find the distance of the trajectories from the synchronization manifold, and repeat for a parameter that adjusts stability, which we take as $\sigma$.  
The MSF is validated if SE$(t)$ remains small when criterion (\ref{eq:MSF_criterion}) is satisfied.

\begin{figure}[tb!] %%% fig 2 
\centering
    \includegraphics[width=1\columnwidth]{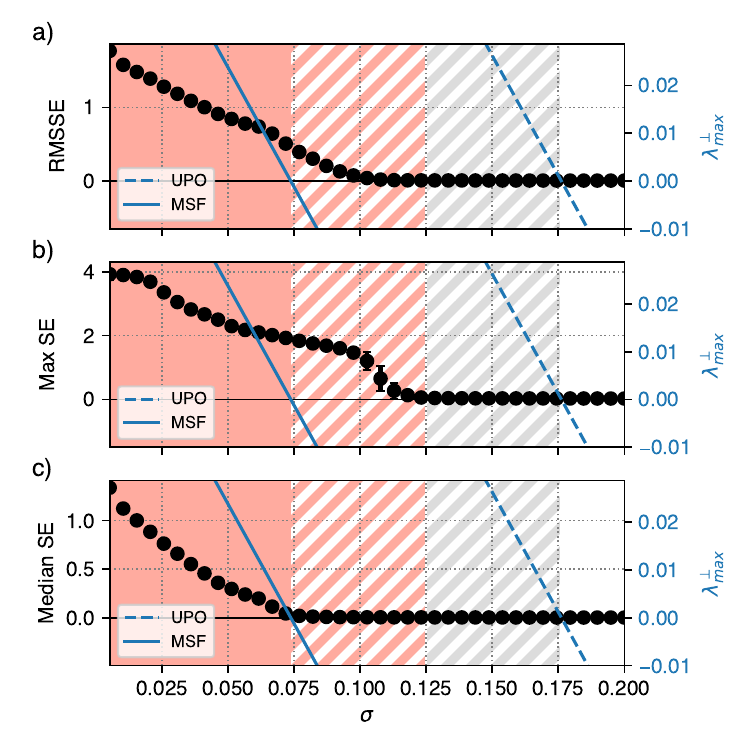}
    \caption{\textbf{The domain of bubbling in a random network of 50 Rössler oscillators.}  The a) maximum, b) mean, and c) mean values of the synchronization error (left axis) as a function of the coupling parameter. Note the different scales of the left vertical scale in panels a) and c) relative to b). The right vertical axis shows the largest transverse Lyapunov exponent (labeled MSF, solid blue line) and the largest transverse Lyapunov exponent of the most unstable periodic orbit (labeled UPO, dashed blue line). The red region is where synchronization is unstable according to the MSF analysis ($\lambda_{max}^\perp > 0$) and the hatched area indicates the region where there is at least one unstable periodic orbit. This area is divided into a red-hatched region, where we observe bubbling, and a gray region, where we do not. The model equations and parameters are as described in {\it Methods}. 
 } 
    \label{fig:5}
\end{figure}

The MSF is based on the asymptotic stability of the synchronization manifold, and hence the root-mean-square SE ({RMSSE}) must approach 0 in the long-time limit. Figure \ref{fig:5}a) shows the RMSSE (left axis, black) and $\lambda_{max}^{\perp}$ (right axis, solid blue line) as a function of $\sigma$. The threshold coupling strength $\sigma_{th} \simeq 0.075$ can be read off from this plot (where $\lambda_{max}^{\perp}=0$).  However, we see that the RMSSE approaches zero at a slightly higher value, indicating that the MSF does not correctly predict the domain of synchronization.

Bubbling can be detected by comparing the maximum value of the synchronization error, denoted by Max SE, to the RMSSE.  Figure \ref{fig:5}b) shows that Max SE stays high for considerably larger values of $\sigma$, and only approaches 0 for $\sigma \approx 0.125$ above which there is bubble-free, high-quality synchronization. Unfortunately, studies of network synchronization usually measure the RMSSE (see, for example, \cite{Sun2009,Pecora2013,Nazerian2023,Bayani2024}). 

Here, we propose a new metric to clearly detect bubbling based on comparing the median SE (Median SE) and Max SE. The Median SE, shown in Fig. \ref{fig:5}c), is insensitive to bubbling, and we see that it approaches zero when $\sigma=\sigma_{th}$, as predicted by the MSF. Therefore, the bubbling region can be delineated as the region where the Median SE $\approx 0$ while Max SE remains significantly above 0 (the red hatched region in Fig. \ref{fig:5}).

\textbf{Criterion \#2: Transversely Unstable Periodic Orbits:} Previous studies of bubbling demonstrate that it can arise from transversely unstable invariant sets embedded in the synchronization manifold.  Thus, bubble-free behavior is predicted when
\begin{equation}
    \boxed{~~\lambda_{max}^{\perp,UPO}<0,~~} \label{eq:criterion_UPO}
\end{equation}
where $max$ indicates the largest transverse Lyapunov exponent for all UPOs, which typically has a low period \cite{Heagy1995}. 

The most transversely unstable UPO for the network has an associated exponent indicated by the blue dashed line in Fig.~\ref{fig:5}.  We see that it becomes negative above $\sigma \sim 0.175$.  Thus, we expect that bubbling occurs for $\sigma_{th}<\sigma<0.175$ However, we do not observe any bubbling in the gray-hatched region in Fig.~\ref{fig:5}, where there are transversely unstable UPOs, but Max SE and Mean SE are very small.

Therefore, we conclude this criterion is too restrictive for networks of R\"{o}ssler oscillators; it is a sufficient but not necessary condition.

\begin{figure}[tb] %%%fig 3
\centering
    \includegraphics[width=\linewidth]{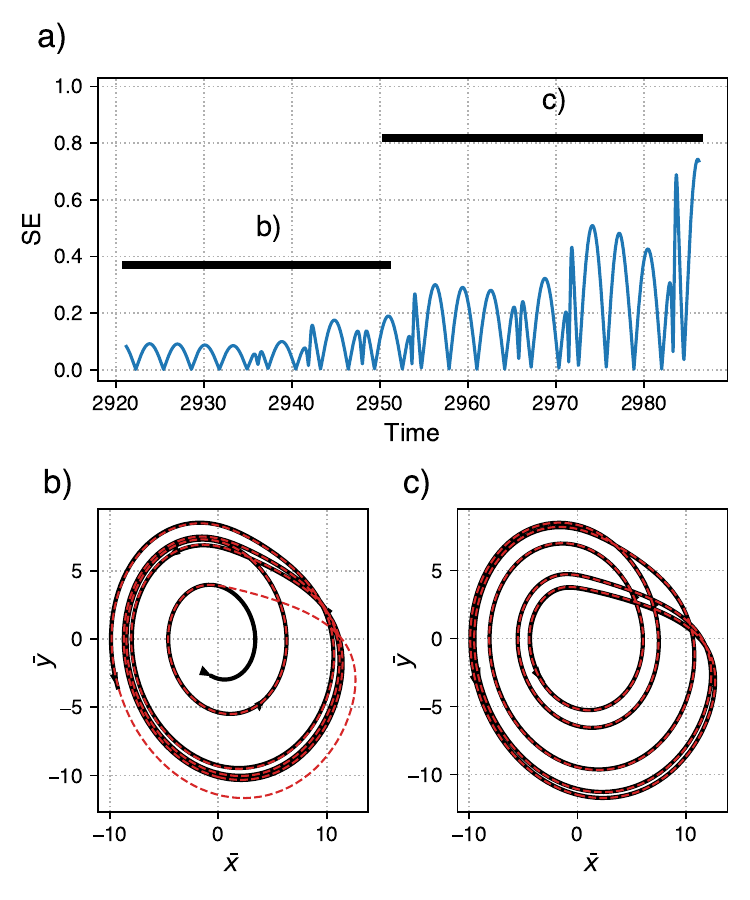}
    \caption{\textbf{Destabilizing synchronization by an unstable periodic orbit.} Development of the bubbling event presented in Fig. \ref{fig:storyboard}. a) Evolution of the synchronization error. b) Average trajectory (black line) projected onto the $(x, y)$ plane during the first time interval marked in a) and a period-5 UPO (dashed red line). c) Average trajectory (black) during the second interval marked in a) and another period-5 UPO (dashed red line).}
    \label{fig:2}
\end{figure}

\textbf{Criterion \#3: Cascaded UPOs:}  Bubble formation has two conditions: the trajectories on the synchronization manifold must be near a transversely unstable UPO ($\lambda_j^{\perp,UPO}>0$) and they need to remain near this UPO long enough for the bubble to grow.  The residence time is approximately given by $\tau_j^{UPO}$ defined in Eq.~(\ref{eq:UPO_time}).  We find that the trajectories approach transversely unstable UPOs, but do not remain long enough to generate a bubble when $\sigma$ is in the gray-hatched region in Fig.~\ref{fig:5}.

Consistent with a previous study \cite{Heagy1995}, we find that the trajectory can approach a transversely unstable UPO followed by an approach to another transversely unstable UPO.  We call this \textit{cascaded amplification} as it increases the probability of a bubble by extending the time that the trajectories remain in a transversely unstable region of the synchronization manifold.

\begin{figure}[htb] %%fig 4
    \centering
    \includegraphics[width=0.8\columnwidth]{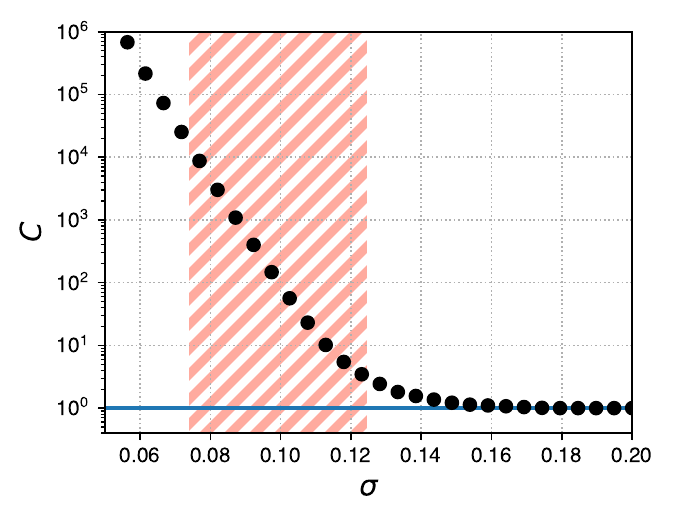}
    \caption{\textbf{Cascaded amplification induces bubbles.}  The cascade amplification factor, $C$ (log scale) quantifies the error amplification produced by the sequence of transitions between transversely unstable UPOs, as a function of the coupling parameter, $\sigma$. The network and the parameters are as in Fig.~\ref{fig:storyboard}. Here, $C=1$ for $\sigma \ge 0.175$ because there are no transversely unstable UPOs above this coupling strength (see the blue dashed line in Fig.~\ref{fig:5}). As in Fig.~\ref{fig:5}, the red hatched area indicates the region where synchronization is unstable according to the MSF analysis, and there is at least one transversely unstable UPO.}
    \label{fig:cascade}
\end{figure}

\begin{figure*}[htb] %%% fig 5
\centering
    \includegraphics[width=\textwidth]{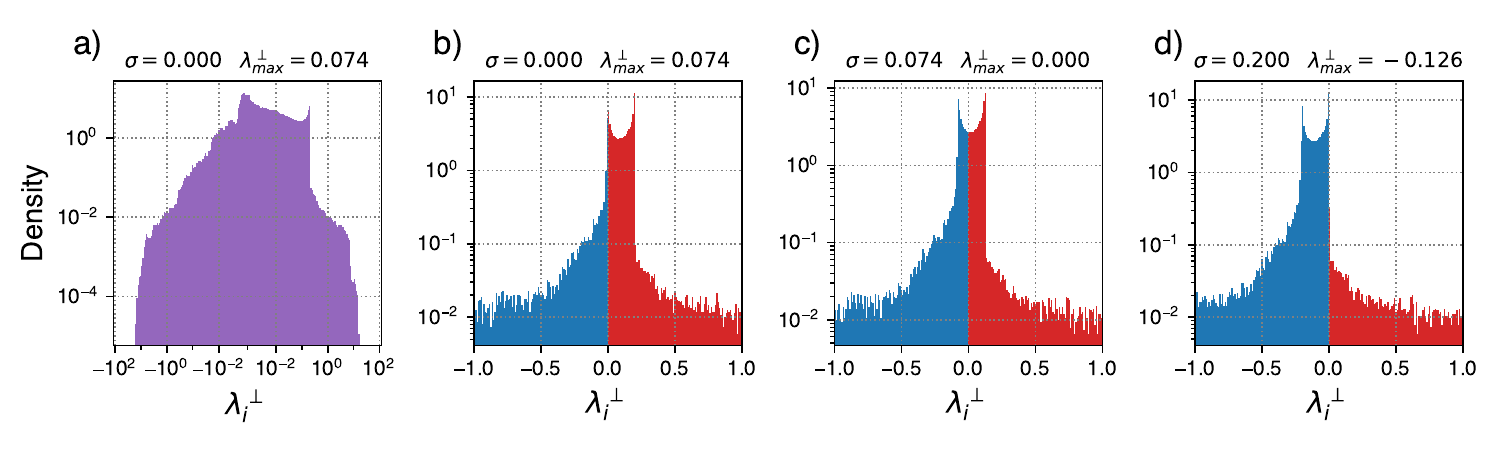}
    \caption{\textbf{Indicators of transverse instability.}  Distribution of the transverse local Lyapunov exponents for no coupling (a), (b) and for values of the coupling parameter c) at d) and above the MSF synchronization threshold. 
    The negative (positive) exponents are shown in blue (red).  
    The average of the distribution, which is the largest transverse Lyapunov exponent,  is also indicated in each panel.}
    \label{fig:3}
\end{figure*}

Figure~\ref{fig:2}a) presents an example of cascaded amplification, showing the temporal evolution of SE at the start of a bubble. Initially, the SE is small, indicating a well-synchronized state, and the trajectories, shown in Fig.~\ref{fig:2}b), evolve near a transversely unstable period-5 UPO.  Near the end of the evolution near this orbit, the SE approximately doubles compared to its initial value.  The trajectories leave the neighborhood of this orbit and then quickly transition to the vicinity of another transversely unstable UPO shown in Fig.~\ref{fig:2}(c). The bubble continues to grow, and, at the end of this second stage, the SE$(t)$ has reached a value that is approximately 10 times higher than its initial value, as seen in Fig.~\ref{fig:2}(a).  

When $\sigma$ is not too large, there are many transversely unstable UPOs and the probability of cascaded amplification is high.  As $\sigma$ increases, more UPOs become transversely stable and, therefore, fewer transversely unstable orbits are available to produce cascaded amplification.

To quantify this relation, we define the cascaded amplification factor that describes the worst-case bubble that can be generated, which occurs when the trajectory only visits transversely unstable UPOs, and visits all of them (\textit{i.e.}, the `perfect storm' scenario).  In this case, the cascade amplification factor is given by
\begin{equation}
    C = \exp\left\{\sum_i \tau_i^{UPO} \lambda_i^{\perp,UPO}\right\}.
\end{equation}
Here the sum is over all transversely unstable UPOs, and hence $C$ is in the range $1\leq C<\infty$, with $C=1$ when there are no transversely unstable UPOs. 

Figure~\ref{fig:cascade} shows that $C$ gradually decreases with $\sigma$ as expected.  The curve is smooth and has no particular feature to indicate instability.  Once criterion (\ref{eq:criterion_UPO}) is satisfied, $C=1$. Thus, a criterion for bubble-free synchronization likely takes the form
\begin{equation}
    \boxed{~~C< C_{th}~~} \label{eq:C_criterion}
\end{equation}
with $C_{th}$ certainly greater than 1.

Some amplification of transverse perturbations can be tolerated as long as the trajectories remain in a linear neighborhood of the synchronization manifold. To predict $C_{th}$ requires a nonlinear stability analysis and will depend on the specific characteristic of the oscillator dynamics.  

Here, we take an empirical approach by finding $C$ at the bubble - bubble-free transition occurring at $\sigma \sim 0.125$.  At this boundary, $C_{th}\sim 3$, which is plausible.  A normal form nonlinear stability analysis may be useful for predicting $C_{th}$ \cite{Ott_pre_1996, Venkataramani1997}; we leave this to a future study.

\textbf{Criterion \#4: Local Lyapunov Exponents}: Another proposed criterion for high-quality synchronization \cite{Gauthier1996, Pecora1998, Yanchuck2001} is given by 
\begin{equation}
    \boxed{~~\lambda_i^{\perp} < 0~~~\mathrm{for~all}~i.} \label{eq:LLE_criterion}
\end{equation}
That is, every point on the synchronization manifold must be transversely stable. As mentioned above, we expect that there will be substantial variation of the LLEs on the synchronization manifold \cite{Abarbanel1991, Abarbanel1992}.

Figure~\ref{fig:3} shows the probability density for $\lambda_i^{\perp}$ for different coupling parameter values. For $\sigma=0$ [panels a) and b), note the log-log scale in a)], the distribution of transverse LLEs is identical to the distribution of longitudinal LLEs and is sharply peaked at a positive value. As expected, the average of this distribution is positive because each uncoupled oscillator is chaotic.   

For $\sigma=\sigma_{th}$ (Fig.~\ref{fig:3}c)), which is the MSF synchronization threshold given in Eq.~(\ref{eq:MSF_criterion}), the distribution is similar to the uncoupled case, but is shifted toward negative values.  The average of the distribution is zero as expected because $\sigma=\sigma_{th}$ is the MSF synchronization boundary. However, we see a long tail of positive LLEs, which can generate bubbling.

For much higher coupling shown in panel d), the average of the distribution moves to negative values, but a long tail of large positive transverse LLEs remain. While the positive exponents are a potential source of transverse instability, we do not observe bubbling for this coupling value. Therefore, we conclude that criterion (\ref{eq:LLE_criterion}) is overly conservative: it is a sufficient but not a necessary condition. 

\textbf{Criterion \#5: Finite-time transverse Lyapunov exponents:}  To generate a bubble, the oscillators' trajectories must reside in a transversely unstable part of the synchronization manifold for a long enough time to allow the instability to grow above the baseline SE. We do not know the regions of the attractor that have positive LLEs or how long the trajectory spends in these regions. 

\begin{figure}[tb!] %%% fig.6
    \centering
    \includegraphics[width=\columnwidth]{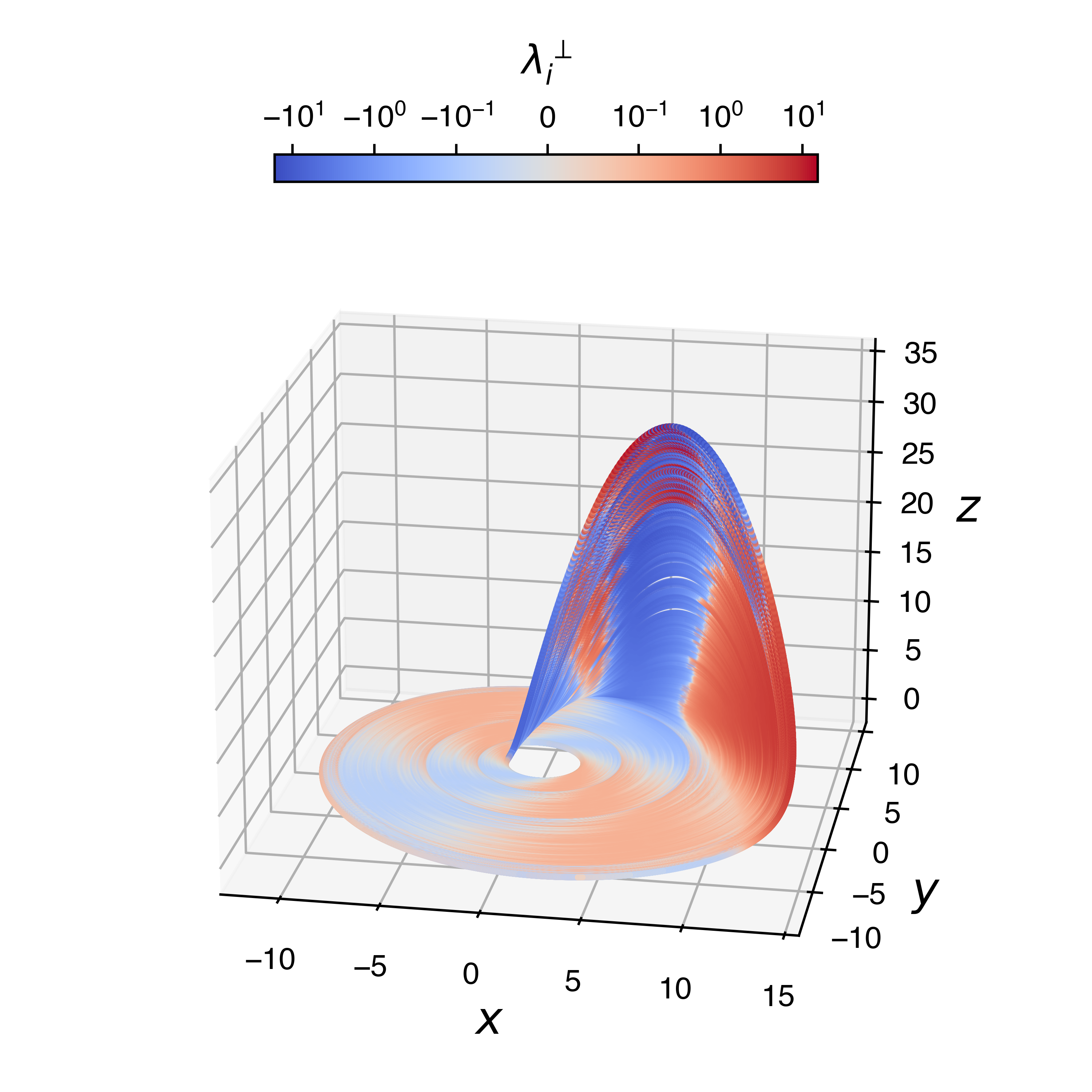}    
    \caption{\textbf{Visualizing the indicators of instability.}  The color bar indicates the value of the transverse local Lyapunov exponents $\lambda^\perp_i$ for $\sigma=0.075$.}
    \label{fig:attractor}
\end{figure}

To shed light on the distribution of LLEs along the attractor, Fig.~\ref{fig:attractor} shows a typical trajectory on the attractor (\textit{i.e.}, on the synchronization manifold). Each point is colored according to the value of $\lambda_i^{\perp}$, where red (blue) indicates transversely unstable (stable) regions.  We observe that most of the attractor is composed of large regions where positive and negative LLEs alternate.  Considering that a bubble develops over several cycles of the attractor (see Fig.~\ref{fig:5}) and that the trajectory passes multiple times through stable and unstable regions, the time spent in the unstable regions must be, on average, longer than the time spent in the stable regions to create a bubble.  A similar heterogenous distribution of local Lyapunov exponents is found for a network of Lorenz oscillators as shown in Supplementary Fig.~1.

To quantify this effect, we define an amplification factor $\mathcal{A}_m$, which varies around the attractor. It is defined in terms of the finite-time Lyapunov exponents $\bar{\lambda}_{m}^{\perp}$ for a time interval $\tau_m$ (see Eq.~(\ref{eq:finite_time_def})) and is given by
\begin{equation}
    \mathcal{A}_m=\exp(\bar{\lambda}_{m}^{\perp}\tau_{m}),
\end{equation}
To find $\mathcal{A}_m$, we search for the interval $\tau_{\mathcal{A}^{max}}$ that achieves the largest amplification $\mathcal{A}^{max}$, which occurs at one region of  the synchronization manifold. 

\begin{figure}[tb] %% fig. 7
    \centering
    \includegraphics[width=0.8\linewidth]{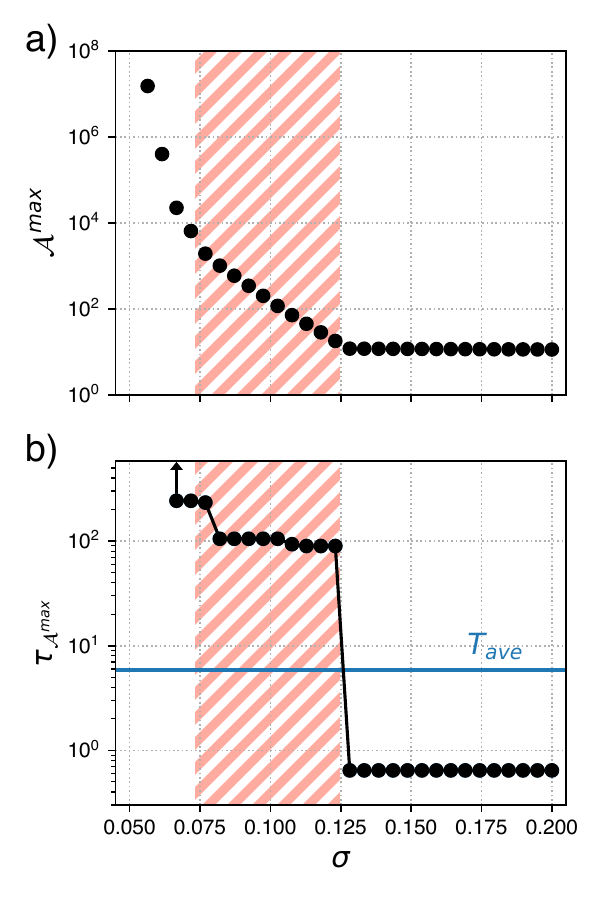}
    \caption{\textbf{Finite-time transverse Lyapunov exponents identify bubbling region.} Finite-time amplification factor, $\mathcal{A}_{max}$, as a function of $\sigma$ (a) and corresponding window size of maximum amplification, $\tau_{\mathcal{A}_{max}}$, as a function of $\sigma$ (b). The average oscillators' pseudo-period, $T_{ave}$, is reported as a horizontal blue line. As in Figs.~\ref{fig:5} and \ref{fig:cascade}, the red-hatched area indicates the region where synchronization is unstable according to the MSF analysis, and there is at least one transversely unstable UPO.}
    \label{fig:finitetimegain}
\end{figure}

Figure~\ref{fig:finitetimegain} shows $\mathcal{A}^{max}$ and $\tau_{\mathcal{A}^{max}}$ as a function of the coupling parameter.  Below the MSF threshold ($\sigma<\sigma_{th}$), $\mathcal{A}^{max}$ is very large ($>10^3$) and $\tau_{\mathcal{A}^{max}}$ tends to $\infty$. This means there is positive amplification during most time intervals, consistent with the MSF prediction of transverse instability. 

Close but below the MSF threshold ($\sigma=\sigma_{th}$), there is an abrupt change in the slope of the curve and $\tau_{\mathcal{A}^{max}}$ goes from $\infty$ to $\approx$ 250.  For $\sigma$ slightly larger than $\sigma_{th}$, $\mathcal{A}^{max}$ decreases exponentially with $\sigma$, and $\tau_{\mathcal{A}^{max}}$ becomes nearly constant at a value of $\approx $100, which corresponds to $\approx$17 cycles around the attractor.  Here, the average period $T_{ave}= 5.9$, which is estimated by averaging 100 time intervals between consecutive crossings of the $x=0$ plane. In this coupling region, $\mathcal{A}^{max}$ is in the range of $10^1-10^3$, which is consistent with the transverse instability needed for bubbling.

At $\sigma \approx 0.125$, which is where we the bubble - bubble-free transition (\textit{i.e.}, the coupling parameter above which no bubbling is observed), $\mathcal{A}^{max}$ changes slope again to a value $\approx 0$, and $\tau_{\mathcal{A}^{max}}$ jumps to a small value ($\approx$ 0.6).  This corresponds to 10\% of the average period around the attractor, which is about the time that the trajectory spends in the most unstable part of the attractor (the blue vertical region in the right part of the attractor in Fig.~\ref{fig:attractor}). 

For $\sigma>0.125$, the amplification is still large ($\mathcal{A}^{max}\sim 10^1$), which explains the periodic increase and decrease of SE in the early part of Fig.~\ref{fig:2}a): The SE is low when the trajectories are in a stable region of the attractor, and grow when they are in an unstable region. However, the large growth needed for a fully developed bubble is prevented because the trajectories do not spend long enough in the unstable region before moving to the stable region of the attractor.

Therefore, Fig.~\ref{fig:finitetimegain} demonstrates that the finite-time transverse Lyapunov exponents, which determine $\mathcal{A}^{max}$ and $\tau_{\mathcal{A}^{max}}$, govern bubbling.  The changes in the slope of $\mathcal{A}^{max}$ and jumps in $\tau_{\mathcal{A}^{max}}$ are distinct and delineate the regions of instability, bubbling, and high-quality synchronization.  

Given the abrupt drop in $\tau_{\mathcal{A}^{max}}$ to a value less than the average period around the attractor, we propose a new criterion for bubble-free synchronization given as 
\begin{equation}
    \boxed{~~~\tau_{\mathcal{A}^{max}} < T_{th}.~~} \label{eq:FTLE_criterion}
\end{equation}
Here, $T_{th}$ depends on the oscillator dynamics.  For our study of chaotic R\"{o}ssler oscillators, a good criterion is obtained for $T_{th}=T_{ave}$. As discussed above, this result makes physical sense since the initiation of a burst takes place over several cycles around the attractor.

To assess the generality of this criterion for other types of networked oscillators, in Supplementary Fig.~2, we show $\mathcal{A}^{max}$ and $\tau_{\mathcal{A}^{max}}$ for a network of chaotic Lorenz oscillators~\cite{lorenz} and find a jump to small values in $\tau_{\mathcal{A}^{max}}$ like that seen in Fig.~\ref{fig:finitetimegain}b).  However, $\tau_{\mathcal{A}^{max}}<T_{ave}$ before the jump and $dt$ after. Thus, $T_{th} < (1+\varepsilon)dt$, where $\varepsilon$ is a small positive number.  We conjecture that this difference is due to a bubble being initiated when the Lorentz trajectories visit a neighborhood of an unstable steady state (with zero period) embedded in the synchronization manifold.

\begin{figure}[tb] %% Fig. 8
    \centering
    \includegraphics[width=0.5\textwidth]{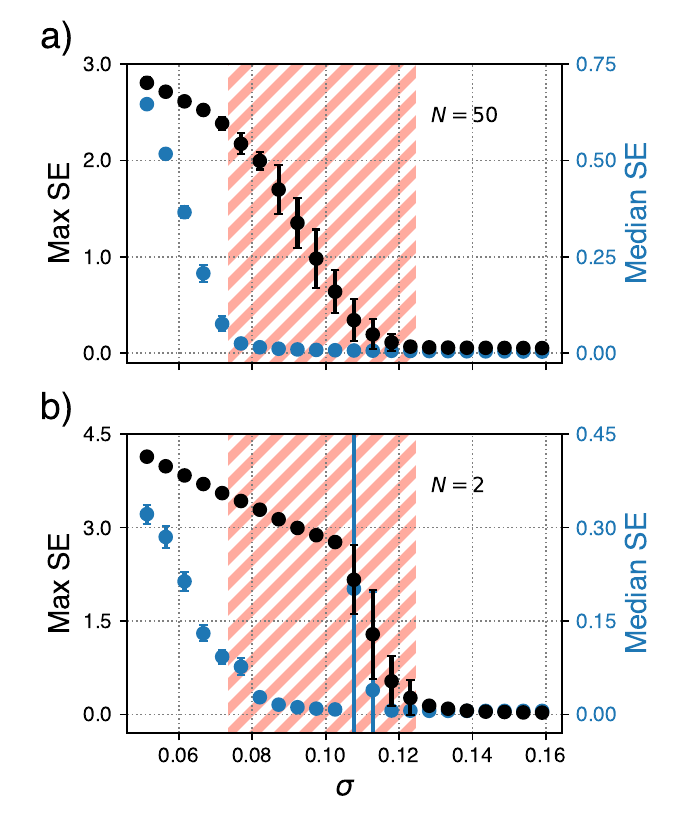}
    \caption{\textbf{Bubbling is largely unaffected by the network size.} Comparison between the maximum and median synchronization error of a network of 50 Rössler oscillators (a) and of 2 Rössler oscillators (b). 
    The heterogeneity is a) $\pm0.5\%$ and (b) $\pm0.18\%$, such that the heterogeneity level is similar to the $N=50$ case. Note the difference in the scales of the left (Max SE) and right (Median SE) vertical axes.} 
    \label{fig:Net_size}
\end{figure}

\textbf{Dependence of bubbling on network size:}  
The previous results are obtained for a single fixed network size and topology. In this section, we demonstrate that the domain of bubbling is largely unaffected by the network size as long as we properly scale the oscillators' heterogeneity. To observe bubbling, smaller networks need to be less heterogeneous than larger networks because of the correlation between heterogeneity and the most unstable transverse direction. For networks with $\sim 5$ or more oscillators, the correlations average out and the heterogeneity level required for bubbling becomes independent of the network size. See \textit{Methods} for the scaling of the heterogeneity level with the network size. 

Figure~\ref{fig:Net_size} compares Median SE and Max SE as a function of the coupling parameter for a larger network ($N=50$) and for two coupled oscillators.  The vertical dashed line indicates the MSF threshold ($\sigma_{th}\simeq0.075$).  We see that the Median SE approaches zero for $\sigma>\sigma_{th}$ near $\sigma \sim\sigma_{th}$ for the larger network, while it occurs for somewhat higher coupling strength for two coupled oscillators. In both cases, the Max SE remains high well beyond the MSF threshold, indicating bubbling, and approaches zero at about the same value of $\sigma$ for both network sizes. The results are similar to the results shown in Fig.~\ref{fig:5} for an intermediate network size. Thus, we argue that the bubbling phenomenon occurs in oscillator networks with random topology regardless of the network size. 

  \begin{figure}[tb] %% Fig. 9
    \centering
    \includegraphics[width=0.5\textwidth]{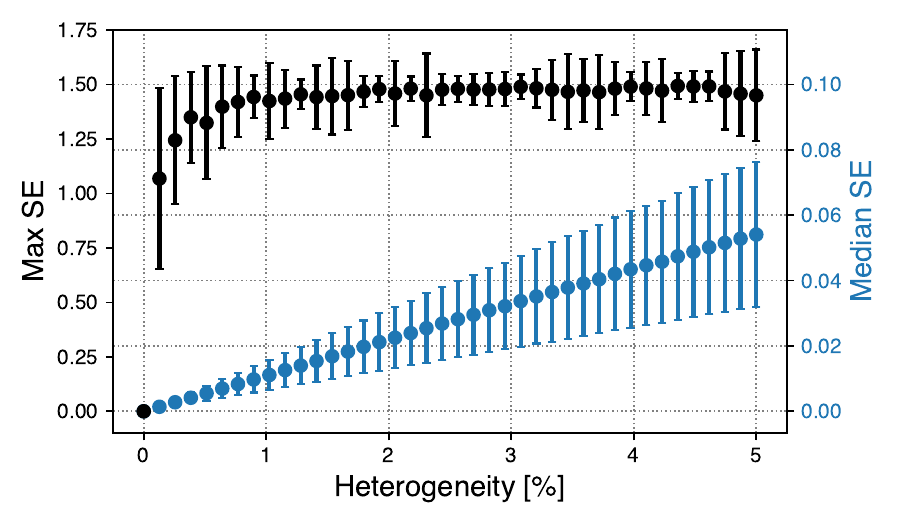}
    \caption{\textbf{Bubbling is largely unaffected by the heterogeneity level.} Comparison between the maximum and median synchronization error for the network of Fig.\ref{fig:storyboard} for different values of the parameters' heterogeneity and a value of the coupling parameter, $\sigma=0.1$, within the bubbling region %(cf. Fig. \ref{fig:5})
    .}
    \label{fig:heterog}
\end{figure}

\textbf{Network Heterogeneity:} 
The results described above are obtained with small node heterogeneity ($<0.5$\%). Here, we show that bubbling persists when there are larger parameter mismatches.
Figure \ref{fig:heterog} shows, the Max SE and Median SE as a function of the level of heterogeneity for $\sigma = 0.1$. We see that Median SE increases linearly as previously reported \cite{Sun2009}, but the Max SE first increases rapidly and then saturates. Bubbling appears even when the level of heterogeneity is small ($\sim$0.1\%), which is difficult to achieve in real networks with a large number of oscillators. However, when the heterogeneity is too large ($\sim \pm$5\%), Max SE is only one standard deviation larger than Median SE, which hinders bubbling detection.

\textbf{Bubbling in cluster synchronization: }  For coupling strengths smaller than those needed for full network synchronization, it is possible to observe that a few oscillators in a `cluster' synchronize their behavior.  The MSF formalism has been extended to predict the coupling thresholds at which the different clusters synchronize \cite{Sorrentino2016, Schaub2016, Bayani2024}.  These clusters are associated with what are known as external equitable partitions for the diffusive coupling scheme considered here \cite{Cardoso2007,OClery2013}.  Based on our results above, we expect that cluster synchronization will be disrupted by bubbling.

\begin{figure*}[tb] %%fig 10
    \centering
    \includegraphics[width=\linewidth]{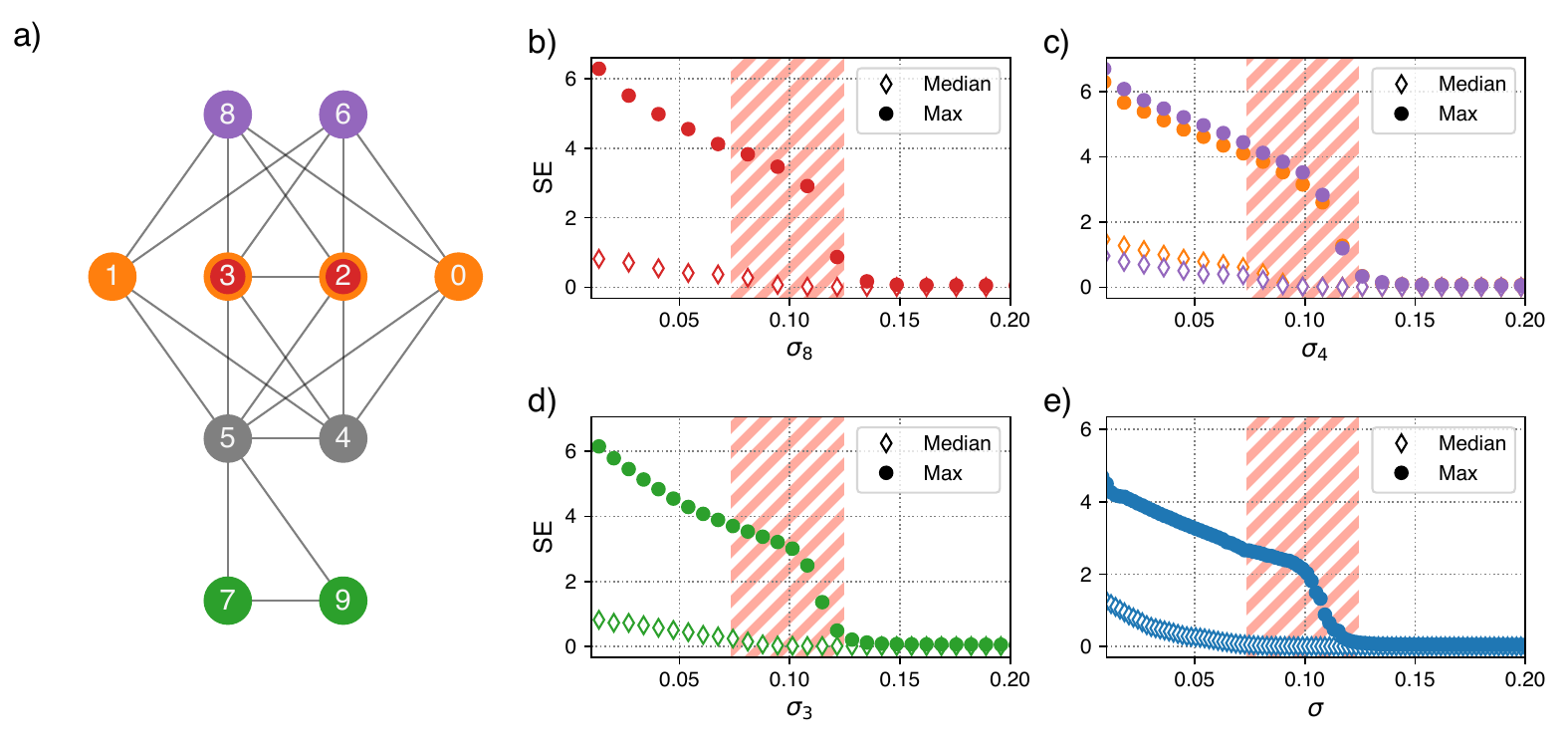}
    \caption{\textbf{Bubbling in cluster synchronization.} a) Network of 10 Rössler oscillators displaying cluster synchronization, where the possible clusters are marked by different colors; %Nodes 0 and 1 form the orange cluster merging with the red cluster, already formed by nodes 2 and 3. 
    The Median and Max SE as a function of the coupling parameter, $\sigma_r = K\gamma_r$ for each cluster, with curves color-coded accordingly to the clusters in panel a). b) Cluster (2,3) with $\gamma_8= 6$, c) clusters (6,8) and (0,1) with $\gamma_4=4$, d) cluster (7,9) with $\gamma_3 = 3$, and e) the entire network with $\gamma_2 = 0.898$. {The heterogeneitity is $\pm$0.5\%.}}
    \label{fig:cluster}
\end{figure*}

To explore this question, we designed a small network with $N=10$, depicted in Fig.~\ref{fig:cluster} (a), where the link symmetries create external equitable partitions \cite{Cardoso2007, OClery2013, Schaub2016}. The MSF predicts that the four clusters formed by these partitions will become stable for different values of coupling strength $K$ (Supplementary Fig.~3). The threshold for synchronizing each cluster depends on the threshold for complete synchronization and the eigenvalue $\gamma_\zeta$ from the Laplacian matrix associated with each cluster (see Eq.~\eqref{eq:K_cluster} in {Methods}) \cite{Bayani2024}.

The first predicted synchronization transition as a function of $K$ occurs for the red pair (node numbers 2,3), followed by the purple pair (6,8), which coincides with the formation of an orange cluster (0,1,2,3) because these two clusters share the same eigenvalue. Finally, the green pair (7,9) synchronizes. Nodes 4 and 5 synchronize for a larger coupling strength, which also corresponds to the coupling strength needed for complete network synchronization. 
 
Figures~\ref{fig:cluster} b) - d) show the Median and Max SE as functions of the coupling parameter $\sigma_r$, which is related to the eigenvalue associated with each cluster.  Scaling $K$ by the respective eigenvalues results in the thresholds for all clusters being the same (see Eq.~\eqref{eq:sigma_r} and Table~\ref{tab:yourlabel} in {Methods} for details). The hatched area indicates the range between the threshold predicted by the MSF and the value $\sigma_r$ where the Max SE drops to near zero. We notice that bubbling is observed in the transition to high-quality synchronization of every cluster in the network. For the sake of completeness, we show in panel e) the results for the complete synchronization, where $\sigma$ coincides to the same coupling parameters used previously.

Notably, the coupling required to suppress bubbling for the red cluster is about twice the value predicted by the MSF. Moreover, unlike in the case of the entire network, the median SE of the clusters does not agree with the MSF predictions because the coupling strengths required to reduce the median SE of the clusters to near zero are systematically higher than the MSF predictions. Therefore, we conclude that bubbling strongly affects cluster synchronization. 

Once again, we observe an agreement between the end of the bubbling region and the prediction given by the finite-time transverse Lyapunov Exponents (Fig. \ref{fig:finitetimegain}), demonstrating the robustness of this measure in assessing the persistence of bubbling in cluster-synchronized states 
%\textcolor{red}{cris: I don't understand this sentence}.

\section{Discussion}

We demonstrate that brief and large-scale desynchronization events are prevalent in networks of heterogeneous R\"{o}ssler oscillators for coupling strengths where high-quality synchronization is expected according to the MSF analysis. We remark that the MSF analysis is based on a linear stability analysis and is only strictly valid for identical, noise-free oscillator networks.  %or noise.  This behavior is known as bubbling.  

We also show that it is possible to predict the region of coupling strengths where bubbling occurs. There are many criteria for predicting the domain of synchronization and bubbling, and we have made a systematic comparison. The MSF criterion (Criterion \#1) fails to predict bubbling, but its converse predicts the absence of high-quality synchronization. %synchronized behavior.  We advocate that
Therefore, we propose to use
\begin{equation}
    \lambda_{max}^{\perp}>0
\end{equation}
%should be used 
to predict the domain of transverse \textit{instability}.

%We also investigate methods for explicitly predicting the bubbling domain.  
Criterion \#2 (Eq.~(\ref{eq:criterion_UPO})) predicts explicitly the bubbling domain by considering the transverse Lyapunov exponents associated with UPOs embedded in the attractor on the synchronization manifold.  However,
%.  Bubbling is expected when
%\begin{equation}
%   \lambda_{max}^{\perp,UPO}<0~~~~\mathrm{with}~~~~ \lambda^{\perp}<0.
%\end{equation}
%We find that this prediction  
we conclude that this criterion to be overly conservative because we find bubble-free synchronization where Criterion \#2 predicts bubbling.  %It is not a necessary condition.

We introduced Criterion \#3 to account for the time that the system resides in a neighborhood of unstable UPOs. We introduced the cascaded amplification factor $C$, which accounts for the system visiting one transversely unstable UPO after another.  %It assumes that the trajectories visit every unstable UPO and hence is a `perfect storm' scenario.   is 
%\begin{equation}
%  C < C_{th},
%\end{equation}
%where 
While predicting an appropriate threshold value $C_{th}$ likely requires a nonlinear stability analysis, %.  We leave this extended analysis to future work, but 
we conjecture that this criterion is also overly conservative because it assumes that the trajectories visit every unstable UPO and hence is a `perfect storm' scenario.

%it is less than 10 for a typical low-dimensional chaotic oscillator.

Criterion \#4 considers the local transverse Lyapunov exponents and predicts bubble-free behavior when
\begin{equation}
    \lambda_i^{\perp} < 0~~~\mathrm{for~all}~i%~~~~\mathrm{with}~~~~ \lambda^{\perp}<0. \label{eq:LLE_criterion_recap}
\end{equation}
Very large values of the coupling parameter $\sigma$ ($\sim$2) are required to satisfy this criterion because of the long tail in the distribution of $\lambda_i^{\perp}$ (see Fig.~\ref{fig:5}a), yet we find bubble-free synchronization within this domain. Thus, this criterion is even more conservative than Criterion \#3 and is not a necessary condition.  However, it is the safest case for a network engineer requiring stability.

We hypothesize that Criteria \#2 and \#4 overestimate the bubbling domain because they do not consider the time that the system resides in an unstable region of the synchronization manifold.  We therefore introduce Criterion \#5, which is based on the finite-time transverse Lyapunov exponents.  By averaging over a finite-time window, we capture both the strength of the instability and the residence time.

We introduce an amplification factor $\mathcal{A}^{max}$ and the averaging time $\tau_{\mathcal{A}^{max}}$ that achieves this maximum.  We find that $\mathcal{A}^{max}$ shows distinct scaling changes with $\sigma$ at the bubbling boundaries, and that there is a sharp transition in $\tau_{\mathcal{A}^{max}}$ at the bubbling - bubble-free boundary. This leads us to the conjecture that bubble-free synchronization is attained when
\begin{equation}
  \tau_{\mathcal{A}^{max}}<T_{th},
\end{equation}
where $T_{th}=T_{ave}$ is the average period of the R\"{o}ssler oscillator.

Our systematic evaluation of the synchronization criteria is for a network of 50 randomly connected R\"{o}ssler oscillators. We also show that bubbling occurs for different network sizes for similar coupling parameter values $\sigma$.  We also study a network that displays cluster synchronization.  Again, comparing Median SE and Max SE we identify the bubbling region and show that the domain of cluster synchronization is smaller than expected based on the MSF analysis.  This leads us to conclude that bubbling is prevalent in networks of chaotic R\"{o}ssler oscillators. 

We now discuss the limitations of our study.  First, we point out that we did not consider the case of a ring of oscillators, where a long-wavelength spatial instability is possible for high coupling strengths \cite{Heagy1994,Matias1997}.  Based on our analysis, bubbling should also be present in such networks, but further testing is needed to determine whether high-quality bubble-free synchronization can be observed in such networks. 

Second, we necessarily study bubbling over a finite observation time. This results in Max SE, $C$, and $\mathcal{A}^{max}$, and $\tau_{\mathcal{A}^{max}}$ being lower bounds for their true values. Hence, the coupling strength needed to observe bubble-free synchronization may be higher than the values presented here. However, we remark that performing extremely long simulations with constant parameters is unrealistic because slow parameter drifts in real-world systems are unavoidable and, therefore, the system's long-term asymptotic behavior can be expected to be different from the real dynamics (in the same way that identical oscillators are an idealization that can fail to represent real-world systems). 

Third, we focus our study on networks of R\"{o}ssler oscillators; further work is needed to test the universality of our criterion to networks of other types of oscillators. On the other hand, many synchronization studies have considered R\"{o}ssler oscillators because they are popular `toy models' to represent the deterministic but irregular oscillatory behavior often observed in real-world systems. In the Supplementary Material, we present the analysis of a network of Lorenz oscillators, where we find similarities with the R\"{o}ssler oscillators, but also differences that are attributed to an unstable fixed point that the synchronized trajectory often approaches.

Finally, our most promising criterion is based on the finite-time Lyapunov exponents, which are not invariant under a change in metric. As a result, the criterion is not universal.  However, these exponents are nearly invariant, with the deviation from universality decreasing as the averaging time increases \cite{Abarbanel1992}.  In the bubbling region, the averaging time spans tens of typical oscillations of the R\"{o}ssler oscillator, meaning any deviation should be minimal.

Our work has important implications for the experimental study or design of oscillator networks because any real network has small heterogeneity or noise, and the existing criteria either do not work or are overly conservative.  In the future, we will explore methods to anticipate bubbles before they happen to give an early warning of a pending extreme event.  Closed-loop control strategies may prevent them, saving the system from extreme events \cite{Cavalcante2013}. Data-driven machine learning tools \cite{Gauthier2021, Shokar2025} may be a promising approach to forecasting and controlling bubbles.

\vspace*{.25in}
\begin{acknowledgments}
R. P. A. acknowledges partial support from by the European Commission under European Union’s Horizon 2020 research and innovation programme Grant Number 101017716; Coordenação de Aperfeiçoamento de Pessoal de Nível Superior–Brasil (CAPES), Finance Code 001.

C.M. acknowledges partial support from the Ministerio de Ciencia, Innovación y Universidades (PID2021-123994NB-C21); Institució Catalana de Recerca i Estudis Avançats (Academia); Agencia de Gesti\'o d'Ajuts Universitaris i de Recerca (AGAUR, FI scholarship and SGR2021).
\end{acknowledgments}

%TC:ignore

\section{Methods}
\subsection{Model}

We consider a network of $N$ nearly identical Rössler oscillators \cite{Rossler1976}. The evolution of the $n$th network node (with $n=1\dots N$) is given by
\begin{eqnarray}
    \dot x_n &=& -y_n - z_n - K \sum_r^N L_{nr} x_r \nonumber  \\
    \dot y_n &=& x_n + a_n y - K \sum_r^N L_{nr} y_r \label{eq:model} \\
    \dot z_n &=& b_n + x_n(z_n - c_n) - K \sum_j^N L_{nr} z_r,\nonumber 
\end{eqnarray}
where the state of each node is given by $\mathbf{x}^n=(x_n,y_n,z_n)$, $K$ is the coupling strength, $\mathbf{L}$ is the Laplacian matrix with components $L_{nr} = \delta_{nr} \sum_r A_{nr} - A_{nr} $, $A$ is a symmetric adjacency matrix, and $\delta_{nr}$ is the Kronecker delta.  Equation~\ref{eq:model} can be written compactly as $\mathbf{\dot{x}}^n = \mathbf{F}(\mathbf{x}^n)$.

Here, we only consider coupling the same variable to the same variables of the other oscillators for simplicity; we do not consider cross-coupling between the variables.  We only consider coupling all R\"{o}sslear variables because this is the only coupling scheme where all LLEs can enter the stable region \cite{Yanchuck2001} even though most researchers studying synchronization in oscillator networks only couple a single variable. 

Unless stated otherwise, $A$ is a random matrix with a link density of 11\%. The values of the parameters $a_n$, $b_n$, and $c_n$ are drawn from uniform distributions with means of 0.2, 0.2, and 7, respectively, and half-widths equal to 0.5\% unless otherwise stated. 

We define $\mathbf{x} = [\mathbf{x}^1, \mathbf{x}^2, ...,\mathbf{x}^N]^T$, $\mathbf{F}(\mathbf{x}) = [\mathbf{F}(\mathbf{x}^1), \mathbf{F}(\mathbf{x}^2), ..., \mathbf{F}(\mathbf{x}^N)]^T$ and $\mathbf{H}(\mathbf{x}) = [\mathbf{H}(\mathbf{x}^1),\mathbf{H}(\mathbf{x}^2), ...,\mathbf{H}(\mathbf{x}^N)]^T$, where $\mathbf{H}$ %: \mathbb{R}^3 \rightarrow \mathbb{R}^3$ 
is a function of each node's variables and the network topology defines the coupling. We also define a vector $\mu = [\mu_1, \dots, \mu_N]^T$ that encodes the parameters of each node, where $\mu_n = [a_n, b_n, c_n]$ represents the parameters of node $n$.

With these definitions, the dynamical evolution of the network is given by 
\begin{equation}
    \mathbf{\dot{x}} = \mathbf{F}(\mathbf{x},\mathbf{\mu}) - K \mathbf{L} \otimes \mathbf{H}(\mathbf{x}),
    \label{first0}
\end{equation}
where $\otimes$ represents the Kronecker product and $\mathbf{H}$ is the coupling matrix given by
\begin{equation}
    \mathbf{H} = 
    \begin{bmatrix}
        1 & 0 & 0 \\
        0 & 1 & 0 \\
        0 & 0 & 1 
    \end{bmatrix}
    = \mathbf{I}^3.
\end{equation}

Equation \eqref{first0} is integrated numerically using the Livermore Solver for Ordinary Differential Equations (LSODE), which automatically detects stiff problems and adjusts the algorithm accordingly, implemented int the \emph{Python} library \emph{Scipy} (version 1.9.3). Trajectories were recorder at a uniform step size $\Delta t=0.2$.  We integrate Eqs.~\ref{eq:model} for a transient time of 2,000 time units, and then integrate for another 100,000 time units to generate the data for our analyses.  

\subsection{Master stability function (MSF)}

Assuming that the oscillators are identical, 
Eq.~(\ref{first0}), becomes
\begin{equation}
    \mathbf{\dot{x}} = \mathbf{F}(\mathbf{x}) - K \mathbf{L} \otimes \mathbf{H}(\mathbf{x}).
    \label{first}
\end{equation}
The synchronization manifold $\mathbf{x}^\parallel$ is defined by the $N-1$ constraints
\begin{equation}
    \mathbf{x}^1 = \mathbf{x}^2 = ... = \mathbf{x}^N \ 
\end{equation} 
with
\begin{equation}
    \mathbf{L} \otimes \mathbf{H}(\mathbf{x}^\parallel) = \mathbf{0} \ ,
\end{equation}
which leads to the completely synchronized solution $\mathbf{\dot{x}}^\parallel = \mathbf{F}(\mathbf{x}^\parallel)$.

To study the stability of the synchronization solution, we analyze how small perturbations $\delta\mathbf{x}^n = \mathbf{x}^n - \mathbf{x}^\parallel$ around it evolve, where $\mathbf{x}^\parallel = [x_\parallel, y_\parallel, z_\parallel]^T$. Applying linear stability analysis for each node, we have
\begin{equation}
    \mathbf{\dot{x}}^n \approx \mathbf{F}(\mathbf{x}^\parallel) + D\mathbf{F}(\mathbf{x}) \Big|_{\mathbf{x}^\parallel} \delta\mathbf{x}^n - K \sum_r L_{nr} D \mathbf{H}(\mathbf{x})\Big|_{\mathbf{x}^\parallel} \delta\mathbf{x}^r,
\end{equation}
where $D\mathbf{F}$ is the Jacobian matrix. % that represents the differential of $\mathbf{F}$ at every point where $\mathbf{F}$ is differentiable. %\textcolor{red}{This does not seem to be correct - DF is the Jacobian, or DH, but not D alone?}. 

The dynamics of the perturbation is then given by
\begin{eqnarray}
    \delta\mathbf{\dot{x}}^n &=& \mathbf{\dot{x}}^n - \mathbf{\dot{x}}^\parallel = D\mathbf{F}(\mathbf{x}) \Big|_{\mathbf{x}^\parallel} \delta\mathbf{x}^n \nonumber \\ &-& K \sum_r L_{nr} D \mathbf{H}(\mathbf{x})\Big|_{\mathbf{x}^\parallel} \delta\mathbf{x}^r,
    \label{eq:noname}
\end{eqnarray}
%Then, we can rewrite Eq.(\ref{eq:noname}) as
which we can rewrite as
\begin{equation}
    \delta\mathbf{\dot{x}} = [\mathbf{I}_N \otimes D \mathbf{F}(\mathbf{x}^\parallel) - K \mathbf{L} \otimes D\mathbf{H}(\mathbf{x}^\parallel)] \delta \mathbf{x} \ .
    \label{diag}
\end{equation}

%Now let's turn our attention to the Laplacian matrix.
In the case of symmetric and undirected coupling, the Laplacian matrix has the following properties:
\begin{itemize}
    \item A set of real positive eigenvalues $(\gamma_r \geq 0)$
    \item The associated eigenvectors constitute an orthonormal basis. % of $\mathbb{R}^N$
    \item The smallest eigenvalue is $\gamma_1 = 0$, which is associated with the eigenvector
\end{itemize}
\begin{equation}
    V_1 = \pm\frac{1}{\sqrt{N}}(1,1,1,...,1)^T \ .
\end{equation}
Hence, $V_1$ is aligned with the synchronization manifold $\mathcal{S}$, and the other eigenvalues have associated eigenvectors that span the phase space transverse to $\mathcal{S}$. %cite bocca
Due to these properties, $\mathbf{L}$ is diagonal 
\begin{equation}
    \mathbf{V}^{-1} \mathbf{L} \mathbf{V} = \mathbf{\Gamma} = \text{diag}(\gamma_1, \gamma_2, ..., \gamma_N)
\end{equation}
where $\mathbf{V} = [V_1,V_2,...,V_N]$ is an orthonormal matrix whose columns are eigenvectors of $\mathbf{L}$, and $\mathbf{\Gamma}$ is a diagonal matrix whose diagonal elements are associated eigenvalues ordered by magnitude.

We define a new set of variables $\eta = (\mathbf{V}^{-1} \otimes \mathbf{I}_m)\delta\mathbf{x}$, 
%so that Eq. \eqref{diag} becomes
%\begin{equation}
%    (V \otimes I_m) \dot{\eta} = [\mathbf{I}_N \otimes D \mathbf{F}(\mathbf{x}^s) - K %L \otimes D\mathbf{H}(\mathbf{x}^s)] (V \otimes \mathbf{I}_m)\eta.
%\end{equation}
%solving for $\delta\eta$
%\begin{equation}
%    \dot{\eta} = (V^{-1} \otimes \mathbf{I}_m) [\mathbf{I}_N \otimes D \mathbf{F}%(\mathbf{x}^s) - K L \otimes D\mathbf{H}(\mathbf{x}^s)] (V \otimes \mathbf{I}_m)\eta
%\end{equation}
which gives us
\begin{equation}
    \dot{\eta} = [\mathbf{I}_N \otimes D \mathbf{F}(\mathbf{x}^\parallel) - K \mathbf{\Gamma} \otimes D\mathbf{H}(\mathbf{x}^\parallel)] \eta
    \label{msf} \ .
\end{equation}
Equation \eqref{msf} is the \textit{Master Stability Function} (MSF) \cite{Pecora1998}, a block-diagonalized variational equation where each block has the form
\begin{equation}
    \delta\dot{\eta}_r = [D\mathbf{F}(\mathbf{x}^\parallel) {-} \sigma_r D \mathbf{H}(\mathbf{x}^\parallel)] \delta\eta_r.
    \label{msf3}
\end{equation}
We define a parameter 
\begin{equation}
    \sigma_r = K\gamma_r\,
    \label{eq:sigma_r}
\end{equation}
where $r$ represents the eigenvector associated with an eigenvalue $\gamma_r$ of the matrix $L$. Therefore, the variational equation \eqref{diag}, with $\eta_1$ accounting for the motion along $\mathcal{S}$, is decoupled from the other variables $\eta_r~ (r > 1)$ representing the dynamics transverse to $\mathcal{S}$.

%\raul{Dan: This is not needed - why repeat an entire equation when you can just give the definition of $\sigma$???  We define a parameter $\sigma_r = K\gamma_r$, which allows us to write a linear parametric equation that depends solely on $\sigma_r$ and is given by %, the dynamics of the uncoupled system $D\mathbf{F}$, and the coupling scheme $D\mathbf{H}$}:
\begin{equation}
    \delta\dot{\eta}_r = [D\mathbf{F}(\mathbf{x}^\parallel) {-}  \sigma_r D \mathbf{H}(\mathbf{x}^\parallel)] \delta\eta_r.
    \label{msf3}
\end{equation}
Given the Laplacian $L$ of the network and its eigenvalues $\gamma_r$, we compute the Lyapunov exponent of Eq. (\ref{msf3}), denoted by $\lambda^\perp(\sigma_r)$, and determine whether $\lambda^\perp(K\gamma_r) < 0$ for $r>1$.  We only have to monitor the second smallest eigenvalue ($r=2$) if we assume that the MSF has a unique intercept with the horizontal axis at a critical coupling strength $K^*$ for complete synchronization.  Therefore, all the results for complete synchronization are presented as a function of the coupling parameter defined as 
\begin{equation}
   \sigma = K\gamma_2.
\label{eq:sigma_definition}
\end{equation}
We denote the Lyapunov exponent associated with this direction as the largest transverse Lyapunov exponent $\lambda_{max}^{\perp}$.

The stability of synchronized clusters can be determined by the MSF formalism as discussed in \cite{Pecora2013, Sorrentino2016, Schaub2016, Bayani2024}. Here, we follow the algorithm proposed in \cite{Bayani2024} to analyze the stability of cluster states. First, the threshold for complete synchronization is determined using the MSF, given by \( \sigma^{*} = K^{*} \gamma_2 \), where the complete synchronized state is stable if \( K > K^* \). Next, the threshold $K^s\zeta$ for the stability of a synchronized cluster synchronized state is found by rescaling \( \sigma^{*} \) with respect to the eigenvalue \( \gamma_{\zeta} \) associated with the specific cluster $\zeta$ of interest (see \cite{Bayani2024} for details): 
\begin{equation}
    K^\zeta = \frac{K^*\gamma_2}{\gamma_\zeta} \ .
    \label{eq:K_cluster}
\end{equation}
We then apply the same rescaling for the threshold obtained by the finite-time Lyapunov exponents. Table \ref{tab:yourlabel} shows the eigenvalues and coupling thresholds associated with the cluster stability for the network analyzed in Fig.~\ref{fig:cluster}.
\begingroup
\setlength{\tabcolsep}{10pt}
\begin{table*}[!htbp]
\centering
\begin{tabular}{||l||l||l||l||}
\hline
 Cluster & Eigenvalue $\gamma_\zeta$& Coupling threshold (MSF)         &  Coupling threshold (FTLE) \\ \hline \hline
$(2,3)$        &  $\gamma_{8} = 6$     &  $\sigma^*/\gamma_{8} \approx$ 0.0125  & $\sigma^{FTLE}/\gamma_{8} \approx$  0.0201\\
$(6,8)$        &  $\gamma_{4} = 4$     &  $\sigma^*/\gamma_{4} \approx$ 0.0187 & $\sigma^{FTLE}/\gamma_{4} \approx$  0.031\\
$(0,1,2,3)$    &  $\gamma_{4} = 4$     &  $\sigma^*/\gamma_{4} \approx$ 0.0187 & $\sigma^{FTLE}/\gamma_{4} \approx$  0.031\\
$(7,9)$        &  $\gamma_{3} = 3$     &  $\sigma^*/\gamma_{3} \approx$ 0.025 & $\sigma^{FTLE}/\gamma_{3} \approx$  0.041 \\
(all nodes)  &  $\gamma_{2} = 0.898$ &  $\sigma^*/\gamma_{2} = K* \approx$ 0.0835               &   $\sigma^{FTLE}$ = 0.125\\ 
\hline\hline
\end{tabular}
\caption{\textbf{Thresholds for stable cluster-synchronized states depend on the eigenvalues associated with each cluster}. The threshold for complete synchronization ($K^*\gamma_2$) can be used to estimate the threshold for cluster-synchronized states by rescaling with the eigenvalue associated with each cluster \cite{Bayani2024}. The same rationale applies to the threshold obtained using FTLE ($\sigma^{FTLE}$), which ensures high-quality synchronization.}
\label{tab:yourlabel}
\end{table*}
\endgroup

\subsection{MSF with small heterogeneity}
As shown in \cite{Sun2009}, the MSF can be adjusted to take into account small heterogeneities in the parameters.
\begin{equation}
%    \delta\dot{\eta}_r = [D\mathbf{F}(\mathbf{x}^\parallel) \giulio{-}  \sigma_r D \mathbf{H}(\mathbf{x}^\parallel)] \delta\eta_r + D_\mu \mathbf{F}(\mathbf{x}^\parallel)\sum_{i=1}^N V_r^i\delta\mu_i
    \delta\dot{\eta}_r = [D\mathbf{F}(\mathbf{x}^\parallel) {-}  \sigma_r D \mathbf{H}(\mathbf{x}^\parallel)] \delta\eta_r + D_\mu \mathbf{F}(\mathbf{x}^\parallel)\sum_{i=1}^N v_{ri}\delta\mu_i    
    \label{msf4}
\end{equation}
%\raul{Is this equation 8 of \cite{Sun2009}? If so we should use another letter for $V$. See that your $V_{r}^{i}$ is an element of the matrix that encodes the eigenvalues of a modified Laplacian, see the paragraph after Eq. 6 in \cite{Sun2009}. In the particular case of $N = 2$ the eigenvectors of the Laplacian and the modified Laplacian are the same, so your results are still valid...}
%\giulio{The modification is for directed graphs. In our case (undirected graphs) the two coincide (see also footnote number 3 of the paper)}
where $v_{ri}$ is an element of $\mathbf{V}$, an orthonormal matrix whose columns are eigenvectors of $\mathbf{L}$, the Laplacian matrix. $\delta\mu_i = \mu_i - \bar\mu$ is the heterogeneity of the parameters with respect to the average value $\bar\mu$ {(here we considered undirected networks, see \cite{Sun2009} for details)}. The heterogeneity values $\delta\mu$ are extracted from a uniform random distribution having a width relative to the parameters' mean value, however, the results do not change significantly when choosing another distribution. In particular, we considered a distribution width of $\pm0.5\%$. The term $\sum_{i=1}^N {v_{ri}}\delta\mu_i$ is therefore a random variable having 0 mean and variance $VAR[\sum_{i=1}^N {v_{ri}}\delta\mu_i] = VAR[\delta\mu]$, given that the eigenvectors are normalized. For a uniform distribution ranging from $-\epsilon$ to $\epsilon$, such variance is $\epsilon^2/3$, which means that the inhomogeneous term is $\mathcal{O}(\epsilon/\sqrt{3})$. If the number of oscillators is 2, the inhomogeneous factor is exactly $\sum_{i=1}^2 {v_{2i}}\delta\mu_i = \sqrt{2}\epsilon$, therefore, to compare simulations between large $N$ and $N=2$, we have to rescale the heterogeneity of the two oscillators as $\epsilon \rightarrow \epsilon / \sqrt{6}.$

Finally, if the MSF predicts stability, the homogeneous part of the equation vanishes for large times and $\delta\eta_r\underset{t\rightarrow\infty}{\longrightarrow}\mathcal{O}\left(\sum_{i=1}^N {v_{ri}}\delta\mu_i\right) = \mathcal{O}(\epsilon)$, that is, it is linear with respect to the amount of heterogeneity \cite{Sun2009}, as shown in Fig. \ref{fig:heterog}.

\subsection{Lyapunov exponents}\label{sec:Lyap_exponents}

As we saw in the previous section, the MSF requires knowledge of the Lyapunov exponents associated with perturbations transverse to the synchronized manifold. Here, we briefly review the procedure used to calculate them. %revise the mathematical logic behind the calculation of this measure.
%\textcolor{red}{The notation in this section was inconsistent with what is presented above - should use bold and also F rather than f.  Check that I fixed it correctly.}
Lyapunov exponents are commonly used to assess how perturbations to a trajectory evolve over time.  Consider a trajectory $\dot{\mathbf{x}}_1 = \mathbf{F}(\mathbf{x}_1)$ and a nearby trajectory $\dot{\mathbf{x}}_2 = \mathbf{F}(\mathbf{x}_2)$ with small variation $\delta \mathbf{x} = \mathbf{x}_2 - \mathbf{x}_1$. To estimate $\delta \mathbf{x}(t)$, we linearize around the first trajectory %to get:
%\begin{equation} \delta \dot{x} = J \delta x \end{equation}
%where $J$ is the Jacobian of $f(x)$, $J={d f}/{dx}$.
%In a \textit{linear} approximation, 
and calculate the rate of separation between trajectories, from $t$ to $t+t_i$, as
$|\delta \mathbf{x}(t+t_i)| \approx e^{\lambda_i t_i}|\delta \mathbf{x}(t)|$, which leads to
\begin{equation} 
\lambda_i \approx \frac{1}{t_i} \ln\frac{|\delta \mathbf{x}(t+t_i)|}{|\delta \mathbf{x}(t)|} \ ,
\label{eq:lle}
\end{equation}
which is the {\it local Lyapunov exponent (LLE)} that characterizes the stability against small perturbations in the region of the attractor that the trajectory is visiting. Because we are concerned with perturbation along transverse directions, we only calculate \textit{tranverse} Lyapunov exponents and hence $\lambda_i$ is the largest transverse local Lyapunov exponent (LLE).  Definitions for other Lyapunov exponents is given near the beginning of the \textit{Results} section.

\subsection{Lyapunov Exponents Notation Summary}

\begin{itemize}
\item $\lambda_{max}^{\perp}$: largest transverse Lyapunov exponent that, in the MSF approach, determines the linear asymptotic stability of the synchronized state against transverse perturbations. 
\item $\lambda_i^{\perp}$: transverse local Lyapunov exponent (LLE) in the region of the attractor that is visited at time $t_i$; $\lambda_{max}^{\perp}$  is the average of the distribution of $\lambda_i^{\perp}$ values.
\item $\lambda_i^{\perp}$: transverse local Lyapunov exponent (LLE) in the region of the attractor that is visited at time $t_i$;
\item $\lambda_i^{\parallel}$: LLE that describes the evolution of a small perturbation along the synchronization manifold, in the region of the attractor that is visited at time $t_i$. Without coupling, they are the same as the transverse LLEs.
    \item $\tilde{\lambda}_i^{\parallel,UPO}$: Lyapunov exponent of the $i$th unstable periodic orbit on the synchronization manifold (also known as parallel or longitudinal Lyapunov exponent).
    \item $\lambda_i^{\perp,UPO}$: transverse Lyapunov exponent of the $i^{th}$ unstable period orbit.
    \item $\lambda_{m}^{\perp}$: Finite-time transverse Lyapunov exponent, which is the average of $\lambda_i^{\perp}$ over a window of $m$ discrete time steps, $\lambda_{m}^{\perp}=(1/m) \Sigma_{i=1}^{m}\lambda_i^{\perp}$.
\end{itemize}

\bibliographystyle{unsrt} 
\bibliography{BubblingOscillatorNetworks}
\end{document}